\documentstyle[prl,aps,epsfig]{revtex}

\newcommand{\be}{\begin{equation}}
\newcommand{\ee}{\end{equation}}
\newcommand{\bea}{\begin{eqnarray}}
\newcommand{\eea}{\end{eqnarray}}
\newcommand{\bdis}{\begin{displaymath}}
\newcommand{\edis}{\end{displaymath}}

\newcommand{\rb}{Rayleigh-B\'enard }
\newcommand{\nsrb}{Rayleigh-B\'enard}
\newcommand{\kad}{Castaing et al. }
\newcommand{\SS}{Shraiman-Siggia }
\newcommand{\nskad}{Castaing et al.}
\newcommand{\nsSS}{Shraiman-Siggia}
\newcommand{\SaS}{Shraiman and Siggia }
\newcommand{\nsSaS}{Shraiman and Siggia}
\newcommand{\deluca}{De Luca et al.}

\newcommand{\bi}{\begin{itemize}}
\newcommand{\ei}{\end{itemize}}

\newcommand{\vv}{{{\bf v}({\bf x},t)}}

\newcommand{\ie}{i.e.}
\newcommand{\la}{\left\langle}
\newcommand{\ra}{\right\rangle}

\newcommand{\role}{r\^ole}

\begin{document}
\bibliographystyle{prsty}

\title{On the Heat Transfer in \rb systems}
\author{R. Benzi$^1$, F. Toschi$^{2,3,4}$, R. Tripiccione$^{4}$}

\date{\today}
\address{$^{1}$  Dipartimento di Fisica, Universit\`{a} di Tor Vergata\\
Via della Ricerca Scientifica 1, I-00133 Roma, Italy\\
$^{2}$ Istituto Nazionale di Fisica della Materia, unit\`a di Tor Vergata\\
$^{3}$ Dipartimento di Fisica, Universit\`{a} di Pisa,\\
Piazza Torricelli 2, I-56126, Pisa, Italy\\
$^{4}$ Istituto Nazionale di Fisica Nucleare, sezione di Pisa}

\maketitle

\begin{abstract}
In this paper we discuss some theoretical aspects concerning 
the scaling laws of the Nusselt number versus the Rayleigh number in a \rb cell.
We present a new set of numerical simulations and compare our findings
against the predictions of existing models.
We then propose a new theory which relies on the hypothesis of Bolgiano
scaling. Our approach generalizes the one proposed
by Kadanoff, Libchaber and coworkers and solves some of the
inconsistencies raised in the recent literature.

\end{abstract}

\vskip 0.3 truecm
%PACS. 47-25 - Turbulent flows, convection and heat transfer.

\section{Introduction}

In this paper we discuss the scaling properties of heat transport in 
\rb systems. Although this topic has received a lot of
attention during the last years, still there are controversial interpretations
of experimental results. In 1989, Libchaber, Kadanoff and co-workers \cite{kadanoff} have
shown that a new and unpredicted scaling range is observed in \rb systems at
large enough Rayleigh numbers. They also proposed a physical picture
of the experimental observations, based on the crucial r\^ole played by
``plumes'', the coherent structures present in thermal convection, in heat
transport. This picture was somehow questioned by \nsSaS,
who proposed a rather different theoretical approach. Both models make relatively ad hoc assumptions on qualitative
features of the fluid motion in appropriate sub-domains of a \rb cell.
These assumptions, although reasonable on physical ground, are not explicitly
justified by the still poorly understood dynamics, in a system of rather complex phenomenology.

In this paper, we try to build a more dynamical picture of the problem,
by establishing a link between the
scaling properties of heat transport and other well controlled dynamical
features encountered in convective turbulence.
We start presenting  and discussing a new set of results based on
direct numerical simulation which, in our opinion, are able to shed more
light on the problem. 
We then generalize the model of \kad \cite{kadanoff} with a simple ansatz,
whose basic physical meaning is that the dynamics that control Bolgiano
scaling in convective turbulence also modulates the scaling of the Nusselt
number. Our ansatz leads to well defined predictions that we have
verified numerically (and could be experimentally checked).
The obvious objection that Bolgiano scaling (expected to set in at
scales larger than the Bolgiano length) can hardly play any \role in heat 
transfer (where scales of the order of the boundary layers are obviously
important) can be solved by observing, as already pointed out in \cite{new},
that  quantities usually regarded as globally characterizing the flow (such
as energy or temperature dissipation, Bolgiano length) can still be locally
defined in a convective cell as a function of the vertical coordinate. 
In particular, Bolgiano length decreases sharply in regions close to 
the boundary layer. 

Our paper is organized as follows. In section \ref{section2} we introduce the equation
of motion, the dimensionless parameters describing the problem and
Bolgiano scaling. In section \ref{section3} we review the theoretical models proposed
by \kad and \SaS emphasizing the physical assumptions
underlying the two different approach. In section \ref{section4} we present our
numerical simulations and in section \ref{section5}, the most important part of the present work, we discuss and justify a simple ansatz which generalizes the model proposed in \cite{kadanoff}. In section \ref{section6} we present our conclusions.

\label{section1}

\section{The Problem}
\label{section2}

We consider a fluid in a rectangular cell of horizontal size L and
vertical size H. The fluid is heated from below 
and cooled from above by contact with two heat reservoirs. 
The temperature field $T(x,y,z;t)$ satisfies the boundary conditions:
\be
\label{temp-bc}
\left.T\right|_{z=0} = {\overline T} - {{\Delta T} \over 2} \;\;\;\;\;\;\;  \left.T\right|_{z=H} ={\overline T} + {{\Delta T} \over 2}
\ee

Furthermore the vertical walls are supposed to be adiabatic (\ie\ thermal
exchange through these walls
are supposed to be negligible).
We also assume that the fluid flow satisfies the Boussinesq equation
of motion (see \cite{landau}):

\bea
\label{eqn-bous1}
{{\partial \vv} \over {\partial t}} + \left( \vv \cdot \nabla \right) \vv
&=& - {1 \over \rho} \nabla p + \alpha g \theta \hat z + \nu \Delta \vv \\
\label{eqn-bous2}{{\partial \theta} \over {\partial t}} + \left( \vv \cdot
\nabla \right) \theta &=& \chi \Delta \theta\\
\label{eqn-bous3}\nabla \cdot \vv &=& 0
\eea

where $g$ is the acceleration of gravity, $\alpha$ the volume thermal
expansion coefficient, $\nu$ and $\chi$ 
kinematic and thermal diffusivity, $\theta \equiv T-{\overline T}$.
The Boussinesq equations of motion are a first order approximation of buoyancy effects and are supposed to be accurate if thermal gradients are not too strong.
Furthermore the incompressibility condition is valid if velocities are much 
smaller than sound speed.

The relevant dimensionless parameters in equations \ref{eqn-bous1} and
\ref{eqn-bous2} are the Rayleigh number $Ra$, 
the Prandtl number $Pr$ and the aspect ratio $\Gamma$:

\be
Ra \equiv {{\alpha g \Delta T H^3} \over {\nu \chi}}\; ;\;\;\;\;\;\;Pr
\equiv {\nu \over \chi}\; ;\;\;\;\;\;\;\Gamma \equiv {H \over L} 
\ee

For large enough values of $Ra$, the flow becomes turbulent, \ie\ chaotic
both in time and in space. 
One of the basic issues that we want to discuss in this paper concerns how
much heat is transferred from the bottom 
boundary to the top boundary.
%, due to turbulent fluctuations. 
To this end it is useful to introduce the dimensionless number:
\be
\label{def_nuss}
Nu \equiv {{\la v_z T \ra - \chi {{\partial \la T \ra} \over {\partial z}
}} \over { {\chi \Delta T} \over H  }} 
\equiv {{\la \omega' \theta'\ra} \over {\chi {{\Delta T} \over H}}}
\ee

where $\omega'$ is the turbulent fluctuation of the vertical velocity,
$\theta'$ the (turbulent) temperature 
fluctuation and $\la \dots \ra$ stands for space and time averages. $Nu$ is
called the Nusselt number.  It measures the amount of heat transported by 
turbulent fluctuations with respect to the heat transport due to molecular
motion 
(${{\chi \Delta T} \over {H}}$ is the heat flux due to conduction if
the fluid were at rest).

We can now state our problem in terms of the dimensionless parameters so
far introduced: we want to investigate the functional relationship,

\be
\label{def_nuss_law}
Nu = f(Ra,Pr,\Gamma)
\ee

To our knowledge, no meaningful 
effect has been found experimentally on turbulent heat transport due to the
geometric parameter $\Gamma$. 
Therefore, we shall neglect, hereafter, the dependency on $\Gamma$ in
\ref{def_nuss_law}.
We shall confine most of our discussion to the $Pr=1$ case.

If the fluid were at rest the temperature drop inside the cell would be
linear. 
Because of convection the mean temperature profile differs from the purely
conductive one. 
In particular one find that almost all the temperature drop occurs in the
thermal boundary layers.  
The thickness of the thermal boundary layer, $\lambda$,  can be roughly
defined as the thickness of the layer 
over which the temperature drop is nearly equal to $\Delta T /2$. 
An important relation (experimentally well verified) allows us to connect
$\lambda$ to  the $Nu$ number:

\be
\label{nulambda}
Nu \sim {H \over \lambda}
\ee

We remark that relation \ref{nulambda} becomes rigorously true 
if one defines $\lambda$ as follows:
\be
{1 \over {\lambda}} =  \left. {1 \over  {\Delta T}} {{\partial {\la T \ra }} \over  {\partial z}}\right|_{z=0}
\ee

By means of the relation \ref{nulambda} our problem can be rephrased as the
understanding of the scaling of $\lambda$ versus $Ra$.

A related issue that we want to address is the r\^ole of buoyancy forces effects on the statistical properties of turbulent fluctuations for \rb convection. 
Let us introduce the velocity and temperature difference defined as:

\bea
\delta v_{//}(r) &=& \left[ {\bf v}({\bf x}+{\bf r})-{\bf v}({\bf x})
\right] \cdot \hat r\\
\delta T(r) &=& T(x+r)-T(x)
\eea

Following Yakhot \cite{iacotto} we can write:

\bea
\label{yak1}
\la \delta v_{//}(r)^3\ra &=& - {4 \over 5} \epsilon r + {{2 \alpha g}
\over {r^4}} \int_o^r{r'^4 \la \delta T(r') \delta v_{//}(r')\ra dr'} +
6\nu {\partial \over {\partial r}} \la \delta v(r)^2\ra \\
%\ee
%\be
\label{yak2}
\la \delta v_{//}(r) \delta T(r)^2\ra &=& - {4 \over 3} N r + {2 \over
{r^2}} \int_o^r{y^2 \la \delta T(y) \delta v_z(y)\ra dy {{\partial \theta}
\over {\partial z}}} + 6\chi {\partial \over {\partial r}} \la \delta
T(r)^2\ra
\eea

where $\epsilon$ is the mean rate of energy dissipation,
$\epsilon = {\nu \over 2} \int{ \sum_{i,j} \left( \partial_i v_j
+ \partial_j v_i \right)^2 d^3x}$, and $N$ is the mean rate of temperature dissipation, 
$N = {\chi \over 2} \int{ \sum_{i} \left( \partial_i T \right)^2 d^3x}$.

Equations \ref{yak1} and \ref{yak2} replace the well known ``4/5'' 
Kolmogorov equation for homogeneous and isotropic turbulence. 
These equations  have been derived by assuming that small scale turbulent
fluctuations in \rb system are, 
to first approximation, homogeneous and isotropic (see Yakhot
\cite{iacotto}).
The second term on the right hand side of \ref{yak1} and \ref{yak2} 
represents however a non isotropic, thermally driven, contribution.
Therefore \ref{yak1} and \ref{yak2}, although cannot be proven rigorously,
can be useful as guidelines
for describing the difference, if any, between homogeneous isotropic
turbulence and thermal turbulence.

Neglecting intermittency effects, from \ref{yak2} we have the scaling
relation:
\be
\label{yak3}
\delta v(r) \delta T(r)^2 \simeq N r
\ee

There are two physically interesting limit regimes in equation \ref{yak1}: 
either the first or the second term on the right hand side dominates.
In the latter case we deduce the following balance:

\be
\label{dv3}
\delta v^3(r) \sim \alpha g \delta T(r) \delta v(r) \cdot r
\ee

By using \ref{yak3} we obtain:
\be
\label{dv2}
\delta v^2(r) \sim \alpha g r \delta T(r) \simeq \alpha g r \left( {{Nr}
\over {\delta v(r)}}\right)^{1/2}
\ee

We can equivalently rephrase eqns. \ref{dv3}-\ref{dv2} as scaling laws for the velocity and
temperature increments:
\bea
\label{bs1}
\delta v(r) &\simeq& \left( \alpha g \right)^{2/5} N^{1/5} r^{3/5} \\
\label{bs2}\delta T(r) &\simeq& \left( \alpha g \right)^{-1/5} N^{2/5}
r^{1/5} 
\eea

The scaling property defined by \ref{bs1}-\ref{bs2} is referred to as Bolgiano scaling \cite{bolgiano}. 
Using Bolgiano scaling, we can evaluate the second term of the right
hand side of \ref{yak1}. We obtain:

\be
\label{eq1}
{{2 \alpha g} \over {r^4}} \int_0^r{r'^4 \la \delta T(r') \delta v(r') \ra
dr' \simeq \alpha g \delta T(r) \delta v(r) \cdot r \simeq \left( \alpha
g\right)^{6/5} N^{3/5} r^{9/5}}
\ee

From \ref{eq1} we obtain a consistency condition, requiring that  the second term
of the r.h.s. in \ref{yak1} is larger than $\epsilon r$. This is true for scales  $r>L_B$, where:

\be
\label{lb}
L_B \equiv  {{\epsilon^{5/4}} \over {N^{3/4} \left( \alpha g \right)^{3/2}
}}
\ee

$L_B$ is called the Bolgiano scale. For $r<L_B$, the statistical properties
of thermal turbulence 
should be described by Kolmogorov theory of turbulence with scaling:

\bea
\label{ksv}
\delta v(r) &\simeq& \epsilon^{1/3} r^{1/3}\\
\label{kst}\delta T(r) &\simeq& N^{1/2} \epsilon^{-1/6} r^{1/3}
\eea

We remark that equation \ref{kst} corresponds to the scaling of a 
passive scalar while equation \ref{ksv} is the usual Kolmogorov (1941) prediction. 
Finally let us notice that for $r>L_B$ the first term on the r.h.s. of 
\ref{yak2} is always greater than the second term, which represents a consistency condition.

The dependence of both $N$ and $\epsilon$ on $Nu$ and  $Ra$ can be exactly 
derived from the equation of motion. We have (see reference
\cite{sergio} for details):

\bea
\label{exact1}\la \epsilon \ra  &=&  {{\nu \chi^2} \over H^4} \left( Nu-1
\right) \cdot Ra\\
\label{exact2}\la N \ra  &=& Nu \cdot Ra^2 {{\chi^3 \nu^2} \over {H^8
\left( \alpha g\right)^2}}
\eea

Beside any rigorous derivations,  we can give a physical meaning to
\ref{exact1}, \ref{exact2}
using the following arguments. Consider first the mean rate of temperature
dissipation. Since almost all the temperature gradient ($\Delta T$) is across the thermal boundary layer (of thickness $\lambda$), we can estimate:
\be
N \sim \left( {{\lambda} \over H} \right) \left( {{\Delta T} \over \lambda}\right)^2 \sim Nu \cdot Ra^2.
\ee
On the other hand, by using equation \ref{exact1}, we obtain:
\be
\delta v(L_B) \cdot \delta T(L_B) \simeq Nu \cdot Ra
\ee
which is equivalent to say that the Bolgiano length
$L_B$ can be interpreted as the characteristic scale of the eddies 
which transport heat in a \rb system.

Evidences of Bolgiano scaling \ref{bs1} and \ref{bs2}
have been reported both in 2D \cite{bolgiano2D} and 3D \cite{benzi}
numerical simulations of \rb systems (see \cite{benzi2} for a detailed description including the effects of intermittency).

\section{Review of proposed scaling theories on heat transport}
\label{section3}
In this section we review 
some theoretical models, proposed in the past, to derive the
scaling properties of $Nu$ vs. $Ra$ (for a review see also \cite{siggia}).
Our emphasis is on the physical assumptions underlying the models, rather
than on rigorous derivations from
the equations of motion. 

We first discuss three different arguments leading to the old (and currently experimentally disproved) scaling relation $Nu \sim Ra^{1/3}$. Such
relation was predicted by many authors, among them Malkus \cite{malkus54,malkus63}, Priestley \cite{priestley54}, Howard \cite{howard63,howard66}, Spiegel \cite{spiegel62}. 

The first argument assumes that the boundary layer is marginally stable,
i.e.
that the effective $Ra(\lambda)$ number computed at the thermal boundary
layer thickness, $\lambda$, is equal to a 
critical Rayleigh number $Ra_c$, independent of $Ra$. 
Since by definition, $Ra={\alpha g \Delta T H^3} / {\nu \chi}$, we get
$Ra(\lambda) = Ra \cdot (\lambda /H)^3$ and 
hence $Ra = Ra_c (H/\lambda)^3$. Therefore, $\lambda \sim H \left( {{Ra_c}
\over {Ra}}\right)^{1/3}$. 
Using \ref{nulambda} we finally obtain the result $Nu\sim Ra^{1/3}$.

Another way to reach the same result is the assumption of the
independence of the heat flux 
from the height of the cell, $H$. Supposing a scaling of the form $Nu \sim
Ra^{\gamma}$, 
the Nusselt number is just ${F \over {\chi \Delta T \over H}}  
\sim \left( {{\alpha g \Delta T H^3} \over {\nu \chi}} \right)^{\gamma}$,  $\gamma$ is required to be $1/3$, for the heat flux $F$ be independent of $H$. 
From this argument we understand that every model which leads to a
$1/3$ exponent, does, implicitly, assume a decoupling of the top and bottom boundary layers. 

Finally, we want to point out a rather simple argument proposed in
\cite{iacotto} leading to the same result.  
Let us assume that, close to the thermal boundary layer, the statistical
properties of turbulent fluctuations
are not affected by buoyancy effect. This implies, that we can use the
standard dimensional analysis of Kolmogorov
theory. Let us also assume that  the thermal boundary layer thickness, 
$\lambda$, equals the viscous boundary layer thickness, 
$\eta$, (i.e.  $\lambda \sim \eta$). 
Assuming K41 scaling for velocity differences, $\delta v(r) \sim \left( 
\epsilon r \right)^{1/3}$, from the condition $Re \sim 1 \sim {{\delta
v(\eta) \eta } \over {\nu}}$ 
and from the scaling $\epsilon \sim Nu \cdot Ra$ we obtain $Nu \sim
Ra^{1/3}$. 
We remark that this argument is self-consistent, because  $\epsilon \sim
Ra^{4/3}$, 
$N\sim Ra^{7/3}$ so that $L_B \sim Ra^{-1/12}$, which implies  
$\eta \sim \epsilon^{-1/4} \sim Ra^{-1/3}$ and  $\lambda \sim \eta \le
L_B$.
The above argument can also be derived by estimating the energy dissipation
$\epsilon(\lambda)$ in the thermal boundary
layer by the Kolmogorov relation 
$\epsilon \simeq {u(\lambda)^3 / \lambda}$, where $u(\lambda)^2 \simeq
(\alpha g \Delta T \lambda)$. By
assuming that most of the energy dissipation takes place in the thermal
boundary layer and using \ref{exact2}
we obtain $Nu \sim Ra^{1/3}$.

We observe that in all cases, for scales of order $\lambda$ and $Ra$ large
enough, $\lambda \ll L_B$. 
Thus Bolgiano scaling should not be applied near by the boundary layer.

In an important paper, \kad \cite{kadanoff} reported for the first time clear
evidence that 
\be
\label{duesett}
Nu \sim Ra^{\gamma}; \;\;\;\; \gamma\simeq 0.281 \simeq {2 \over 7}
\ee

\kad (1989) have shown that this scaling law is valid for $Ra \ge 10^6$. 
Further experimental results (Ciliberto \cite{sergio}) have shown that $\gamma \sim 2/7$ even
for lower $Ra$.
The results reported in \kad have motivated many experimental and
theoretical efforts aimed
at understanding the physical mechanism leading to the (unexpected) scaling
\ref{duesett}.
Here we review two rather different theoretical model proposed in \kad \cite{kadanoff} and \nsSS \cite{ssiggia}.

The \kad \cite{kadanoff} theory is based upon the assumption that there exist 
three layers (hereafter referred to as A, B and C) characterized by
different physical properties:

\

A-layer: the thermal boundary layer, near the bottom and top
boundaries, of thickness $\lambda$ and 
temperature differences $\Delta T/2$. In the A-layer, instabilities
generate plumes, of typical size $\lambda$, which are expelled into 
the B-layer.

B-layer: a mixing region, of thickness much greater that $\lambda$
and smaller than the size of the cell $H$. 
In this layer thermal plumes are accelerated due to buoyancy effect. We
shall indicate by $\delta T$ and
$\delta v$ the 
 the characteristic size of temperature and velocity fluctuations,
respectively.
 
C-layer: the central region of the cell of size
comparable with the size of the system. 
Velocity  and temperature fluctuations will be indicated by $u_c$ and
$T_c$ respectively. 
In this layer thermal plumes are advected with almost constant velocity.

\

The physical picture behind this theory is the following. Thermal plumes
are generated in the A-layer, accelerated in the B-layer and advected in the C-layer. The Nusselt number can be estimated at the center of the cell 
(where the heat flux is purely convective) as

\be
\label{kada1}
Nu \sim u_c T_c
\ee

To estimate the velocity fluctuations in the center of the cell 
the only dimensional relation (ignoring thermal diffusivity and
kinematic viscosity) 
is the following

\be
\label{kada3}
u_c \simeq (g \alpha T_c \cdot H)^{1/2} 
\ee

The basic assumption of the theory is that in the B-layer the
characteristic velocity fluctuation are given by
the balance between the buoyancy effect of the plume and the viscous
effect, while
temperature fluctuations are equal to the temperature fluctuations carried
out
by the plumes, namely $\Delta T$. Furthermore, the velocity fluctuations
in the B-layer equal the velocity fluctuation in the central region. Thus,
we obtain:

\be
\label{kada4}
u_c \simeq \delta u \simeq {{g \alpha \Delta T \lambda^2} \over \nu}
\;\;\;\;\; \delta T \simeq \Delta T
\ee

From relation \ref{nulambda}, \ref{kada1}, \ref{kada3} and \ref{kada4} it
follows that $Nu \sim Ra^{2/7}$. 
Let us remark that, due to \ref{kada4},  $\lambda$ can be regarded as the
cutoff-scale of velocity fluctuations.
One of the most important point in the \kad theory is the r\^ole played by
the thermal plumes, which are well identified
coherent structures observed in the chaotic dynamics of thermal turbulence.
Equation \ref{kada4} is based on 
the assumption that coherent plumes do exist and are observed in the
B-layer and set the characteristic velocity and temperature fluctuations 
in \rb systems. 

A major criticism on the \kad model concerns the validity of
equation \ref{kada4} in the mixing layer B. Indeed, as already discussed, 
equation \ref{kada4} implies a balance between viscous dissipation and 
buoyancy force. The buoyancy force, however, should be
relevant only for scales larger than $L_B$, the Bolgiano scale. 
The analysis performed in section \ref{section2}, indicates that, for scales close to  the thermal boundary layer $\lambda$, $L_B$ is much greater than
$\lambda$, which implies that velocity fluctuations cannot
be controlled by the overall strength of the buoyancy force.
This implies that the thickness of the thermal boundary layer
cannot be fixed by the balancing proposed in \nskad

In order to overcome this criticism,
a rather different approach has been proposed by \nsSaS, 
whose theory is based on the relevant dynamical r\^ole
played by the mean flow observed in \rb cells. 
The onset of a mean flow is due to
plumes rising from the unstable
boundary layer. On the other hand, the mean flow generates a viscous
boundary layer which, in turns, control
the thickness of the thermal boundary layer. It is assumed, therefore, that
the thermal boundary layer is  contained in the viscous layer. As we shall
see, the most important assumption in the theory
is that all energy dissipation 
is constrained inside the viscous boundary layer.

The starting point of \SaS theory is that, within the thermal boundary
layer, there exists a balancing
between the mean flow advection of horizontal temperature gradient and
vertical thermal dissipation, namely
\be
\label{ss1}
u {{\partial T} \over {\partial x}} \simeq \chi {{\partial^2 T} \over
{\partial z^2}}
\ee

The velocity profile is supposed to be linear inside the viscous boundary
layer 
($\tau$ is a mean shear which has to be determined self-consistently)

\be
\label{ss2}
u \sim {z \over \tau}
\ee

From \ref{ss1} and \ref{ss2}  we can derive a relation between the thermal
boundary layer thickness ($\lambda$) 
and the unknown parameter $\tau$.
%\be
%\label{ss3}
$${\lambda \over \tau} \cdot {{\delta T} \over L} \simeq \chi {{\delta T}
\over \lambda^2}$$
%\ee

\be
\label{ss4}
\lambda^3 \simeq \chi L \tau
\ee

The viscous layer thickness ($\eta$) can be estimated using the requirement

\be
\label{ss5}
{{{\eta \over \tau} \cdot \eta} \over \nu} \simeq 1
\ee

Using the exact relation $\langle \epsilon \rangle = {\it Nu \cdot Ra}$ and
under the assumption that the relevant part of the energy is dissipated
inside the viscous boundary layer,
we obtain
\be
\label{ss6}
\langle \epsilon \rangle = {\it Nu \cdot Ra} \simeq {\nu {1 \over \tau^2}}
{\eta \over H}
\ee

Finally from \ref{ss4}, \ref{ss5} and \ref{ss6} we obtain $Nu \sim
Ra^{2/7}$.

We want to point out that the two theories are based on two
quite different physical
pictures of the basic mechanisms which control heat transport in \rb
system. In the theory by  \kad buoyancy
effects and their balance with dissipation control the characteristic size
of temperature and velocity fluctuations.
According to \nsSaS, on the other hand, buoyancy is responsible to maintain the mean flow
in the system which dynamically
controls temperature and velocity fluctuations. 

In the next section we present further experimental and numerical results
which will clarify the physics of the thermal boundary layer.

We close this section remarking that the scaling properties discussed so far, 
are  not supposed to be asymptotic, as shown by Kraichnan \cite{kraichnan} in the early sixties. 
For very large $Ra$ numbers an asymptotic regime should emerge as  can be
understood by the following argument. 
The maximal velocity which can be reached inside the cell can be estimated
as:

\be
U_M \equiv \left( \alpha g \Delta T H \right)^{1/2}
\ee
Since the maximal rate of energy dissipation associated with this velocity
is $U_M^3 / H = \epsilon$, 
it follows that $Nu \sim Ra^{1/2}$. 
We notice that for such a scaling regime the Bolgiano scale would became
$Ra$ independent.
Asymptotic prediction of $1/2$ exponent can also be derived as a rigorous
upper bound of heat transport
(see Doering and Constantin). 

From the experimental point of view, some evidences of transition to the
asymptotic regime have been recently reported \cite{transizione}.

\section{Numerical results}
\label{section4}
An important insight in the physics of thermal convection
can be obtained by direct numerical simulation (DNS) of \rb cells. 
An obvious limitation of DNS is that the available range in $Ra$
is usually much smaller than in laboratory experiments. 
Nevertheless, DNS are extremely useful
in understanding the validity of different physical assumptions and,
occasionally, in checking the predictions on physical quantities which
are difficult to measure in a real experiments. 
In this section we discuss a set of new numerical simulations
aimed at understanding  the correct physics of heat 
transport in \rb.

As already pointed out, numerical simulation are confined to
rather small $Ra$ number with respect to those available in real
experiments. In \cite{kadanoff} the $2/7$ scaling exponent was
observed for $Ra$ larger than $10^6$. 
The same scaling exponent was reported by \deluca \cite{cattaneo}
in 2D numerical simulation of \nsrb.

Here we show that the $2/7$ scaling exponent
is clearly observed, in 3D, even at rather low $Ra$ number.
We have performed a number of numerical simulations 
using a LBE (Lattice Boltzmann Equation) \cite{lbe} 
code on a parallel supercomputer \cite{ape} for a 3D 
\rb cell of aspect ratio 1. 
For a detailed description of the numerical code used we 
refer the reader to \cite{bolgiano2D}. 
For the temperature field, we have imposed adiabatic boundary 
conditions on vertical walls, while top and bottom walls 
have been kept at fixed temperature 
(respectively $-\Delta T /2$ and $\Delta T /2$). 
Velocity boundary conditions are free-slip on vertical walls 
and no-slip (zero velocity) on top/bottom walls. 
All run has been done at $Pr=1$ and  cover a 
range of about one order of magnitude in 
$Ra$ around $Ra\sim 10^6$.

We have performed runs at different $Ra$ 
in order to measure the $Nu(Ra)$ dependence. 
We found a clear  $Nu\sim Ra^{2/7}$  scaling, see Figure [1].

The results shown in Figure [1] together with those reported
by \deluca, clearly indicate that the scaling exponent does
not depend on the dimensionality of the system. This implies
that arguments based on three dimensional Kolmogorov theory
of turbulence must be ruled out. 

As discussed in the previous section, the viscous 
boundary layer near the top and the bottom of the  \rb
cell plays an important r\^ole in the model suggested by \nsSaS.
In order to clarify this point,
we have performed a new set of
numerical simulations, with 
free-slip boundary conditions for the velocity on the top
and bottom walls.
The choice of this kind of boundary conditions 
allowed us to completely remove the velocity boundary layers, 
while keeping a dissipation of kinetic energy inside the bulk of the cell.
In Figure [1] we have also reported the scaling of $Nu$ versus $Ra$ for the free-slip case. A clear scaling is observed, again, with slope close to $2/7$.

The numerical simulations discussed above indicate that
the viscous boundary layer plays a marginal r\^ole
in determing
the scaling exponent of heat transport. 
We want to remark that in the theory proposed by \SaS one of the basic
physical assumptions is that most of the energy
dissipation occurs in the viscous boundary layer. Certainly
this is not the case in the numerical simulation reported in
Figure [1] (free slip boundary condition), although the scaling exponent
of $Nu$ does not change. Our findings, therefore, seem to rule out
the approach proposed by \nsSaS, at least for what concerns the
assumption on the energy dissipation. 

\section{Bolgiano length and non-homogeneous convective cell}
\label{section5}

We have seen in section \ref{section3} that one of the major criticism 
on the theoretical model proposed by \kad concerns the relevance of buoyancy
force in the thermal boundary layer. The criticism is based on the fact that
the Bolgiano scale ${L_B}$ is much larger than ${\lambda}$ for large $Ra$
number.
By using the observed scaling $\lambda \simeq Nu^{-1} \simeq Ra^{2/7}$,
together with equation \ref{exact1}, \ref{exact2} and \ref{lb}
we obtain $L_B \simeq Ra^{-3/28}$. This result show that the ratio
$L_B / \lambda $ increases as $Ra^{5/28}$.
From a physical point of view, the Bolgiano scale 
represents the scale at which
energy is injected in the system as "potential
energy". Buoyancy force converts this energy 
in kinetic energy.

The analysis made in section \ref{section2} was appropriate for a 
homogeneous/isotropic convective cell. However,
\rb convective cell is not homogeneous neither isotropic.
We can slightly generalize the analysis of section \ref{section2} by
assuming that turbulence in the \rb is 
``locally'' homogeneous and isotropic.
It follows that we must
interpret the various quantities 
(as for example the energy dissipation $\epsilon(z)$, the Bolgiano length
$L_B(z)$ and so on) as depending locally on $z$:
the distance from the bottom wall.

Following this approach, we can introduce 
a local (but plane-averaged) Bolgiano length as the following:
\be
\label{lbz}
L_B(z) \equiv  {{\epsilon(z)^{5/4}} \over {N(z)^{3/4} \left( \alpha g
\right)^{3/2} }}
\ee

Using direct numerical simulations, we can obtain the behavior of
$\epsilon(z)$ and $L_B(z)$ in the \nsrb.
In Figure [2] we report, 
the energy dissipation averaged on horizontal planes as a function of $z$. 
As it is evident, energy is not evenly dissipated inside the cell. 
In Figure [3] the typical behavior of the Bolgiano length, 
$L_B(z)$, as obtained from definition \ref{lbz}, is shown. 

We remark that while the Bolgiano length grows inside 
the bulk of the cell 
(in particular in the center of the cell it reaches its maximum, 
nearly equal to the size of the cell itself) 
it is relatively small near the top/bottom boundaries.

This observation solves the apparent 
inconsistency raised in section \ref{section3} where we noticed 
that $\lambda \ll L_B$:
while the global Bolgiano length, $L_B$, 
can be much greater than the thermal boundary layer thickness, 
locally, the Bolgiano length, $L_B(z)$ has its minimum value 
around the boundary layer thickness itself.

The results shown in Figures [2] and [3] suggest that close
to the thermal boundary layer the buoyancy force is the dominant
effect in the system. Thus, velocity and temperature fluctuations
should be described by the Bolgiano scaling \ref{bs1} and \ref{bs2}.
On the other hand, the thermal boundary layer can be interpreted as
the scale at which dissipation becomes relevant with
respect to buoyancy force.
By using \ref{bs1} and \ref{bs2} and balancing the
second term (namely the forcing) 
in equation \ref{yak1} with the third term, we can introduce
the Bolgiano dissipation length $r_B$. After some
algebraic computation we obtain
\be
\label{rb}
r_B \simeq N^{-1/8}
\ee
$r_B$ is equivalent to the Kolmogorov length $\eta$ for turbulent flows
without buoyancy force.
In a \rb cell, temperature dissipation, $N$, is confined in the thermal boundary
layer. Therefore, equation \ref{rb} can be used locally near the thermal
boundary layer assuming that $N$ is the global mean rate of temperature
dissipation.

Looking again at the results shown in Figure [2] and [3], we are tempted
to assume that in \rb turbulence the thickness of the thermal boundary layer
adjusts itself in such a way that it becomes equal to the Bolgiano
dissipation length, i.e.:
\be
\label{lrb}
\lambda \simeq r_B
\ee
Using equation \ref{rb} and \ref{exact2} we immediately obtain $Nu \simeq Ra^{-2/7}$.
Our ansatz \ref{lrb} generalizes the approach proposed by \kad by using
the basic equation \ref{yak1} and \ref{yak2}.

By using the definition of $L_B(z)$ and the Bolgiano
scaling \ref{bs1} and \ref{bs2} we can compute the value of the Bolgiano
length at scale $\lambda$. We estimate the rate of
energy dissipation as
$\epsilon(\lambda) \simeq \frac{\delta v(\lambda)^2}{\lambda^2}$.
Using \ref{exact2} for the estimate of $N$, we finally obtain:
\be
\label{lblam}
L_B(\lambda) \simeq N^{-1/4} \lambda^{-1}
\ee
From equation \ref{lblam} we immediately see that the ansatz \ref{lrb}
implies:
\be
\label{prediction}
L_B(\lambda ) \simeq \lambda
\ee
Equation \ref{prediction} should be considered a prediction of the
theory so far discussed. 
We show in Figure [4] the values of $L_B(\lambda)$ and $\lambda$
as obtained in our numerical simulations. As one can see, the numerical
results are quite consistent with the prediction \ref{prediction}.

Inside the thermal boundary layer, we can assume, following \nsSaS, that
the velocity profile is linear in $z$ and that the advection of the mean
flow balances the thermal dissipation as expressed in equation \ref{ss1}.
By using the same approach suggested by \nsSS, we obtain that the
mean shear inside the thermal boundary layer should be proportional
to $\lambda ^{-3}$. This implies that the mean rate of energy dissipation
$\epsilon_{\mbox{\tiny B.L.}}$,
integrated inside the thermal boundary layer, is proportional to
$\lambda^{-5}$. Therefore we finally obtain:
\be
\label{transiction}
\epsilon_{\mbox{\tiny B.L.}} \sim Ra^{1/7} \epsilon_{{\mbox{\tiny TOT}}}
\ee
Equation \ref{transiction} is consistent with the numerical simulations,
see Figure [5] where we show $\epsilon_{\mbox{\tiny B.L.}}$ plotted against
$\epsilon_{{\mbox{\tiny TOT}}}$ in a log-log plot.
Equation \ref{transiction} puts bound on the range of $Ra$ where the
$2/7$ scaling regime is observed. More precisely, a transition
to the asymptotic scaling predicted by Kraichnan should be
observed for values of $Ra$ such that 
$\epsilon_{\mbox{\tiny B.L.}} \sim  \epsilon_{{\mbox{\tiny TOT}}}$.
According to Figure [5], the critical $Ra$ number for this transition
is predicted at $Ra \sim 10^{11\pm 1}$ for Prandtl number of order $1$.
This prediction is in reasonable agreement with recent experimental
results found by Chavanne et. al. \cite{transizione}.

%----------------------------------------- Conclusioni
\section{Conclusions}
\label{section6}

The most important result shown in this paper is that the observed
scaling of $Nu$ versus $Ra$ in \rb systems can be explained by assuming
that the thickness $\lambda$ of the thermal boundary layer is controlled by (and also approximately equal to) the
Bolgiano dissipation scale $r_B$, i.e. the scale at which buoyancy forces are
balanced by the dissipative effects. In order to justify our assumption
we have introduced a local Bolgiano scale $L_B(z)$ based on the local values of energy and temperature dissipation.
By using direct numerical simulations of \rb turbulence, we have found that $L_B(z)$
is rather small near the thermal boundary layers and becomes equal to
the cell size near the center of cell.
This is a key point in our analysis because it allows to use
Bolgiano scaling in order to compute the dissipation scale.
Previous investigations started with the observation that $L_B$ is much
larger than $\lambda$ and, hence, for $z \simeq \lambda$ Bolgiano scaling
should be ruled out.
Using the assumption that $\lambda \simeq r_B$, we predict
that $L_B(\lambda) \simeq \lambda$. Such a relation has been
verified numerically.
Our findings support and, in some sense, generalize the model proposed
few years ago by the Chicago group \cite{kadanoff}. Let us also remark that
our results agree quite well with the observed Bolgiano scaling in
direct numerical simulations of \rb systems.

We have not discussed the dependence of $Nu$ on the
Prandtl number $Pr$. Both the model proposed by \kad and \SS predicts
a $Pr$ dependency which scales as $Pr^{-2/7}$. Experimentally, for $Pr \le 1$
the scaling of $Nu$ versus $Ra$ seems to be independent of $Pr$, while
for $Pr<1$ it seems that $Nu \simeq Pr^{a}$ with $a$ small and positive.
The whole behavior is however not completely clear and it is still
under investigation (see \cite{verzicco} for very recent results). At any rate, a negative scaling exponent in the
$Pr$ number is ruled out by existing experimental observations.
The model we have proposed in this paper can partially explain
the experimental results. Indeed, the Bolgiano dissipation scale $r_B$
can be a rather complex function of $Pr$ because either kinematic
viscosity or thermal diffusivity can enter in the computation of $r_B$.
Moreover, if one takes into account intermittent fluctuations and
multiscaling effects on the dissipation scale,
the computation of $r_B$ may explain the observed dependence on $Pr$.
This problem deserves more investigations in the future.

\section*{Acknowledgements}
\label{section7}
This paper have been inspired by a lot of discussions done in the
last few years with L. P. Kadanoff, who taught us to "look at sketchy data
and get (hopefully) deep insight".
We would like also to thank Dr. Simone Cioni, Prof. Sergio Ciliberto and  
Dr. Federico Massaioli discussions.

\newpage
\section*{Figure Captions}
\begin{enumerate}
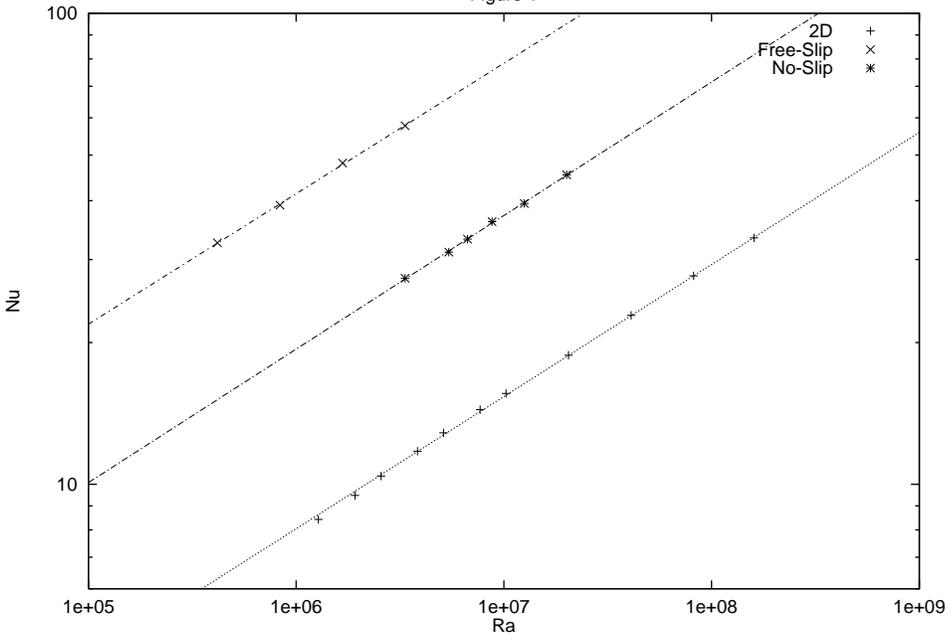
\item [Figure 1] Scaling of $Nu$ versus $Ra$. Data $+$ are taken from \cite{cattaneo} for a 2D numerical simulation (fitted slope $0.280 \pm 0.001$). Data $*$ from our numerical simulations with no-slip boundary conditions (fitted slope $0.283 \pm 0.003$). Same for $\times$ but with free-slip boundary conditions on top/bottom walls (fitted slope $0.277 \pm 0.003$).
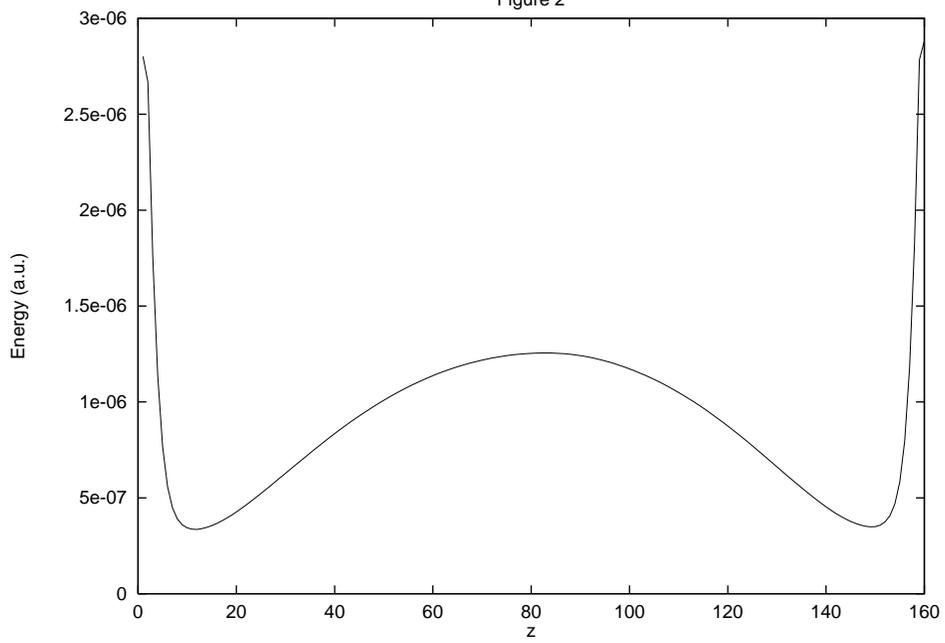
\item [Figure 2] Typical behavior of the energy dissipation (in arbitrary units), $\epsilon(z)$, inside the cell. 
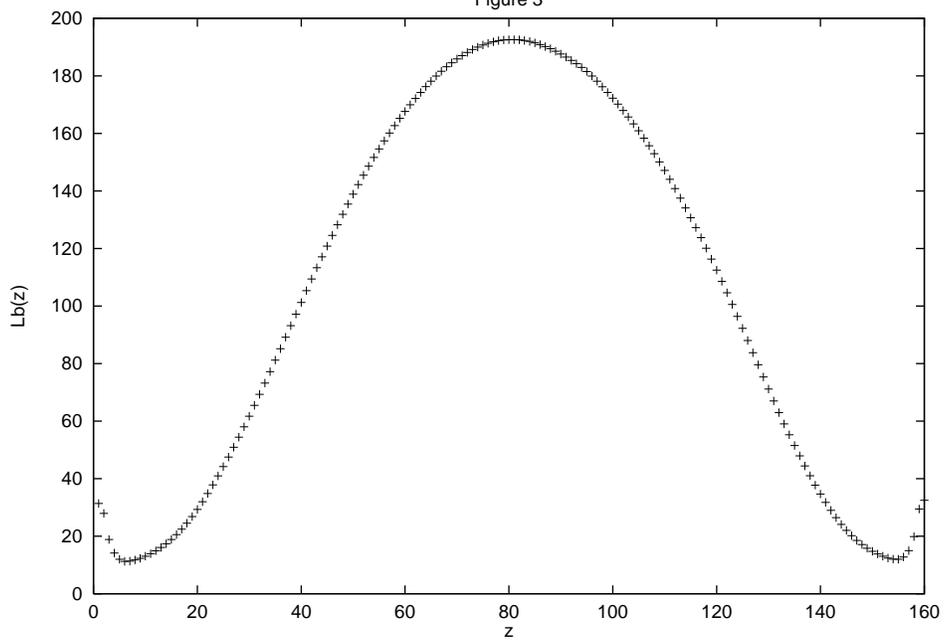
\item [Figure 3] Typical behavior of the Bolgiano length (in lattice units), $L_B(z)$, inside the cell.
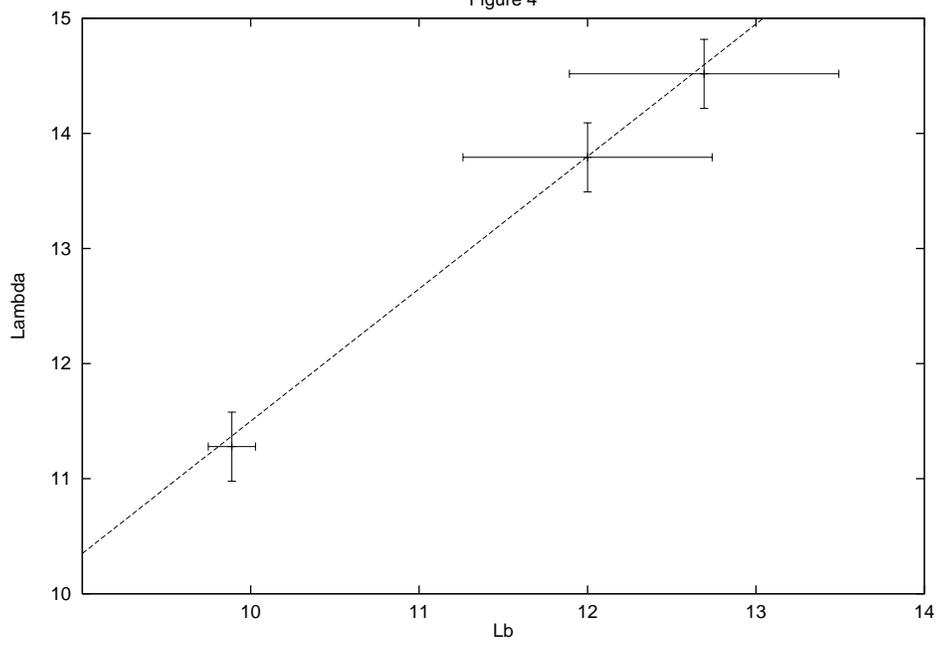
\item [Figure 4] Scaling of $\lambda$ versus $r_B$. The points corresponds to $Ra$ values of $3.34\cdot 10^6,6.68\cdot 10^6,1.67\cdot 10^7$. The straight line is a linear fit, yielding $\lambda = 1.15 \cdot r_B$.
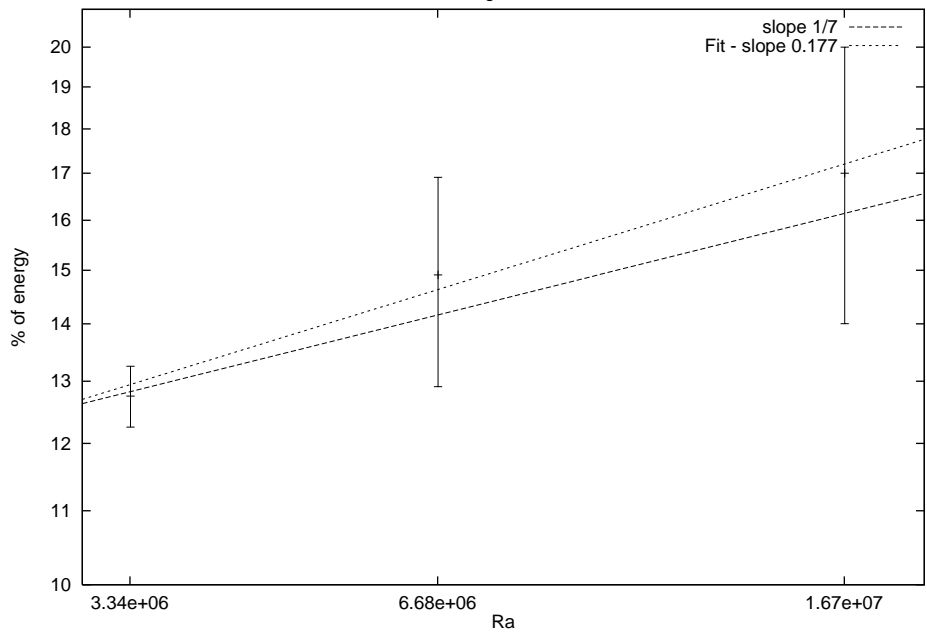
\item [Figure 5] Behavior of the fraction of energy dissipated inside the viscous boundary layers over the total energy dissipate inside the cell. The line are, respectively a fit with the expected slope $1/7$ and with arbitrary slope ($\sim 0.177$).
\end{enumerate}
\end{document}